# The Economic Viability of an In-Home Monitoring System in the context of an Aged Care Setting

**Frank Perri,** Victoria University, Footscray Park, Melbourne, Australia. Email: frank.perri@vu.edu.au

**Shah J Miah**[1]**,** University of Newcastle, NSW, Australia. Email: shah.miah@newcastle.edu.au

**Steve Zanon,** Victoria University, Footscray Park, Melbourne, Australia. Email: steve@proactive-ageing.com

**Keis Ohtsuka,** Victoria University, Footscray Park, Melbourne, Australia. Email: keis.ohtsuka@vu.edu.au

## Abstract

The aged care sector in Australia faces significant challenges. While many of these issues have been clearly identified, their urgency has been further highlighted during the COVID-19 pandemic. Technology such as in-home monitoring is one way to address some of these challenges. However, the efficacy of technology must be considered together with its implementation and running costs to ensure that there is a return on investment, and it is economically viable as a solution. A pilot program was run using the HalleyAssist® in-home monitoring system to test the efficacy of this system. This paper focuses on an economic analysis to better understand the financial viability of such systems.

Using a secondary analysis approach, the findings identified that revenue could be generated by providing carers with additional services such as real-time monitoring of the client, which can foster deeper relationships with the customer, along with savings of healthcare costs to carers, service providers and Government. Savings are related to the earlier intervention of critical events that are identified by the system, as delays in treatment of some critical events can create much more severe and costly health outcomes. Further health costs savings can be made via trend analysis which can show more nuanced health deterioration that is often missed. The implementation of preventative measures via this identification can reduce the chances of critical events occurring which have much higher costs. Overall, monitoring systems lead to a transition from a reactive to a preventative services offering, delivering more targeted and personalised care.

**Keywords:** Aged care, smart home, Australian healthcare, smart technologies, economic-savings

---

[1] Corresponding Author: Prof Shah Miah, Professor and Head of Business Analytics, Newcastle Business School, University of Newcastle, Australia, email: shah.miah@newcastle.edu.au

# The Economic Viability of an In-Home Monitoring System in the context of an Aged Care Setting

**Abstract**

The aged care sector in Australia faces significant challenges. While many of these issues have been clearly identified, their urgency has been further highlighted during the COVID-19 pandemic. Technology such as in-home monitoring is one way to address some of these challenges. However, the efficacy of technology must be considered together with its implementation and running costs to ensure that there is a return on investment, and it is economically viable as a solution. A pilot program was run using the HalleyAssist® in-home monitoring system to test the efficacy of this system. This paper focuses on an economic analysis to better understand the financial viability of such systems.

Using a secondary analysis approach, the findings identified that revenue could be generated by providing carers with additional services such as real-time monitoring of the client, which can foster deeper relationships with the customer, along with savings of healthcare costs to carers, service providers and Government. Savings are related to the earlier intervention of critical events that are identified by the system, as delays in treatment of some critical events can create much more severe and costly health outcomes. Further health costs savings can be made via trend analysis which can show more nuanced health deterioration that is often missed. The implementation of preventative measures via this identification can reduce the chances of critical events occurring which have much higher costs. Overall, monitoring systems lead to a transition from a reactive to a preventative services offering, delivering more targeted and personalised care.

**Keywords:** Aged care, smart home, Australian healthcare, smart technologies, economic-savings

## 1. Introduction

The care of the elderly in Australia faces several challenges; these include a growing ageing population (Piggott, Kendig, & McDonald, 2016), higher numbers of more complex chronic health conditions (Taylor et al., 2019), and an increase in the want of quality services (Gill & Cameron, 2020). These challenges face an increase in care delivery costs. Developing a better understanding of how to improve in-home care while making economic savings in line with the most productive use of limited resources is of high importance. This understanding is currently crucial, as many industry reports, such as those by Stewart Brown, indicate that many aged care service providers are presently unprofitable (StewartBrown, 2020). Along with the financial challenges, aged care service providers must also adhere to ever-stricter quality and regulatory standards while providers compete under the new Consumer Directed Care (CDC) legislation.

Technology and its continued advancement are often seen as part of, if not the solution to many challenges in health care. While technologies can and often do provide solutions, economic viability must also be taken into consideration if the widespread implementation of that technology is to be considered. Studies such as those by Arrieta, Woods, Qiao, and Jay (2014) highlight that even well-established technologies should undergo cost-benefit analyses to ensure viability with their implementation.

A pilot research program[2] was run to trial the benefits of in-home monitoring to assist the provision of client care using the HalleyAssist® smart home system, the findings of this trial, conducted by the National Ageing Research Institute (NARI) are reported separately. While the trial was successful in client acceptance and assisting in providing better quality care, this report focuses on the economic impact of the use of that technology in supporting in-home aged care services.

This study examines the economics of the implementation that technology can provide both as an additional revenue stream and in cost savings. Potential revenue can be generated by providing carers with extra services and information, while savings in healthcare costs could be made on several levels. Both the client and their in-home carers could achieve fewer out-of-pocket expenses from more effective service provision based on collected data analytics. Service providers can also achieve cost savings via a more nuanced allocation of support services from that same data. Government savings could also be achieved from a more effective allocation of support and by providing measurable outcomes to some forms of treatment. Unobtrusive, objective data collection will also assist in future research outcomes, analysis of big data across multiple sites, and public health planning.

## 2. Research Background

Several Government reports and research initiatives have been undertaken to highlight the seriousness of these challenges in aged care. Some of the more notable include the current Royal Commission into Aged Care (Royal Commission into Aged Care Quality and Safety, 2019), The Aged Care Roadmap (Aged Care Sector Committee, 2016) and The National Quality Indicator Program (Australian Government Department of Health, 2019). The effect of COVID-19 on aged care has highlighted that these needs for service improvement must be addressed as a matter of urgency. These reports and investigation outcomes have highlighted the vital role technology will play in the sector, but which forms of technology are practical and economically viable in the longer term are still emerging.

---

[2] The program is based on an industry funded project (Ref- INT – 0780) for assessing impact of Care smart home monitoring technology in the home care setting.

The majority of research in this field has mainly focused on the efficacy of technology in the home or with specific conditions (Amiribesheli & Bouchachia, 2018; Daniel, Irving, Veronica, & Luis, 2020; Lazarou et al., 2016; Vanus et al., 2016). Studies such as these, along with the pilot study to which this analysis is based, measure if the technology is functional, improves health and wellbeing, and accepted by users. In other words, research must ascertain if technology not only works and helps to improve client health but if it is easy to use by the client and their carers. Whilst this type of research is essential to the implementation of technology in the home; economics must also play a role in determining the viability of such systems. A system may work, but it may be too expensive to implement on a broader scale. Technology requires a return on investment so that there is a much larger uptake of that technology. While there has been some research exploring the more general economics of technology in health care (Rahman, Akbar, Rolfe, & Nguyen, 2019), more specific cost analyses with particular forms of technology need to be made. As such, this report focuses specifically on the technology the HalleyAssist® system can provide and the resultant economics. Findings and implications from this study would apply to evaluate other health care technology systems if they offer similar features and benefits in functionality, efficacy, and useability.

## 3. Research Methodology

The study adopted a secondary data analysis of an existing (pilot) project. The pilot project assessed the efficacy of an in-home monitoring system (called HalleyAssist®). A secondary analysis approach is mainly used to reuse existing information to glean new understandings. Secondary analysis is a relatively new approach; it's used to revisit existing data sets or reports that are more widely available online or offline (Tarrant, 2016). Several studies in information systems have formed ways of analysing secondary data in quantitative studies. Examples, such as the study by Campos (2016) used secondary analysis to show existing Facebook advertising data were used to identify the effects of user-generated content. In the same fame, Arnott, Lizama and Song (2017) point out in the context of their analysis of eight business intelligence systems, the increase in data leads to greater generalizability for developing more measures, and is likely to be higher quality when the original researchers are involved. Arnott et al. (2017) suggested that this is due to the original researchers' deep understanding of the data's meaning and further argue that fit between available data and secondary analysis requirement is ensured when similarities in phenomena studied, data collection and unit of analysis apply.

In the current paper, a secondary quantitative (economic) analysis was conducted across the case findings from the pilot study. Costs from a service provider were obtained, along with reported Government and industry averages for financial line items such as hourly rates for wages and on-costs, travel and management expenses. This information was used to examine the economics of the implementation that technology can provide both as an additional revenue stream and in cost savings. This paper aims to assess the economic viability of in-home monitoring systems which will be explored further in the following discussions.

**3.**1 **Case context**

The HalleyAssist® system uses sensors placed in the client's home and artificial intelligence (AI) to monitor movement and map behaviour. This data is analysed for patterns and can be used to notify and assist with support services as well as to alert family members of critical incidents. Sensors such as movement sensors, door sensors and bed sensors are used to collect data which is then analysed via the AI. to identify the individual behavioural trends of each customer. This information can then be used to identify significant behavioural variations, critical events and track declines or emerging risk profiles in day-to-day functioning. Alerts are automatically raised when these situations occur.

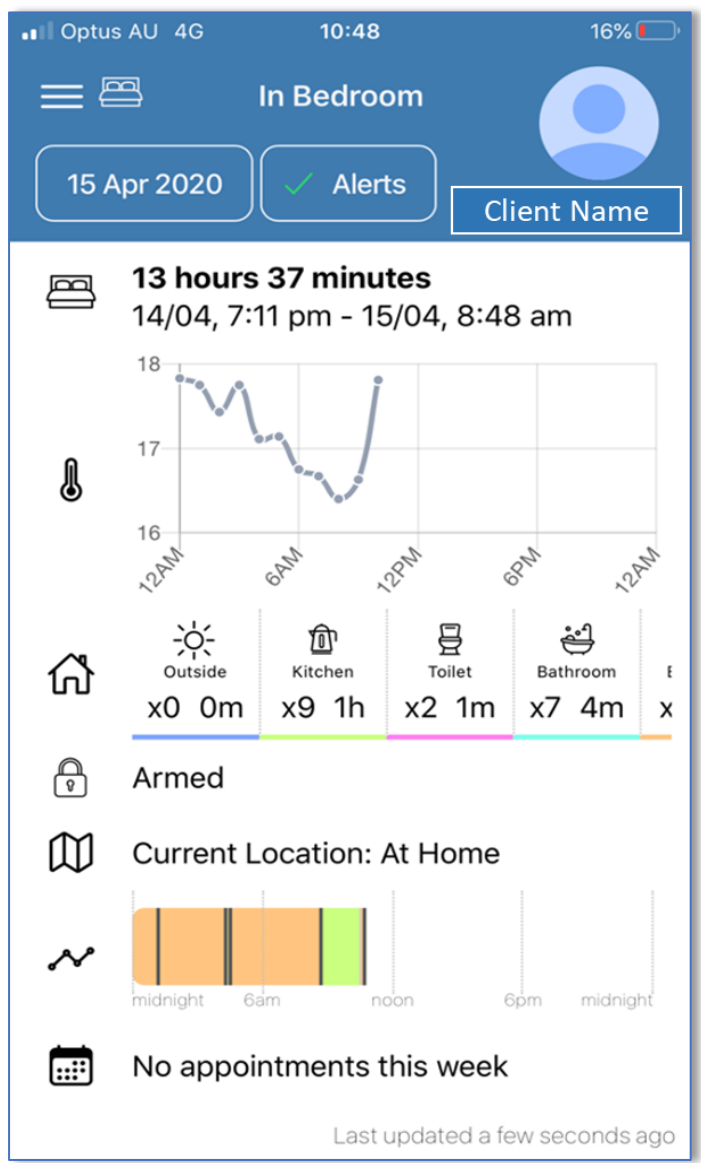

The system includes data dashboards that provide an overview of the client's activities in near-real-time (see Figure 1). The dashboard is accessible to either (or both) the in-home aged care service provider and one or more family/carers so that they can better monitor and gain insights into their loved one's behaviours. Importantly, the system automatically creates alerts if there are issues of unusual behaviour such as falls or a lack of movement. These alerts can then be acted upon by the service provider and or the family. If there are known conditions that are impacting the consumer's quality of life, sensors can be set to track these issues more closely and provide the earliest possible notification of increasing risk so carers can better plan for early interventions.

*Figure 1 : Example of the HalleyAssist® Dashboard*

This pilot program was funded by the Commonwealth Department of Health as part of an Innovation Funding Grant to install the system in several residences. A six-week trial from February to April 2020 of thirteen client homes was undertaken. The trial focused on the acceptability of the technology, measures of the carer burden, the ability to better support clients with specific needs and challenges, well-being, and re-enablement. The trial was found to be successful, with formal research findings reported separately by the National Ageing Research Institute (NARI).

To identify where the primary economic impacts of smart home monitoring systems lie, a detailed financial modelling exercise was undertaken. A series of different scenarios drawn from the trial were examined to offer some 'real-life' examples of the implementation of smart home technology. A sensitivity analysis was undertaken as to how digital technology, in particular, a smart home product could improve the allocation of staff resources and produce financial benefits.

# 4. Findings

During the trial, client care events occurred in which the system was able to monitor and alert researchers. Some of these more notable events became scenarios where cost-benefit analyses could be applied based on if the technology was utilised, versus if the technology was not. Due to the commercial-in-confidence nature of the aged care provider's costs, we cannot divulge specific amounts of savings in these scenarios. Instead, these findings will discuss in broader terms, where extra income can be generated, and savings can be made from the analysis.

**4.1 Family monitoring**

Without a monitoring system, when a potential emergency occurs, family members are often needed to intervene. This checkup activity could include taking time off work and travel expenses. Instead, having information available on a smartphone, where family members and loved ones can observe the status of an individual in real-time would allow them to better identify if an emergency has actually occurred and plan their visit accordingly. In practice, this real-time visibility can minimise carer stress and anxiety, along with the potential to minimise lost productivity.

4.1.1 Day to day monitoring (visibility) for families and carers

Aged care providers could charge for this service by simply providing access to the smart home mobile app for approved consumers at a reasonable rate, thereby increasing revenue. While savings to family members will be situation-specific, some family members will see the benefits of having access to the monitoring system as they will be much less likely to rush out at times from work when there is no emergency.

An additional income stream over and above monitoring fees can be a (phone or home visit) welfare check prompted by either automated alerts from the system or should the family identify a concern from what they might see on the mobile app dashboard. A similar financial model is used in security companies that offer alarm monitoring systems which charge a monthly fee for the monitoring and then an additional fee if the company is requested to go out and visit the location.

4.1.2 Minor events attended to by family

A minor event is defined as an incident where a carer needs to check in on the elderly individual, but they did not require significant assistance—for example, not responding to a telephone call and upon checking finding that they are safe. Dependent on the carers cost of time versus the number of times these incidents occur it may be a financial benefit to the family member to utilise the monitoring system in this way, especially if these events are occurring multiple times. There would be a return on investment for the more 'at-risk' customers, but this may not be needed for every client. This baseline calculation can be used by aged care providers to identify the right clients whose circumstances would warrant a further discussion about the system's installation. There is likely to be a selection (target segment) of clients where this service will be of benefit and the customer would understand the rationale for a return on their investment without them being 'sold to'. Alternately, there will also be those family members that would want the service and monitoring solely for their peace of mind, regardless of how many times callouts are needed.

### 4.1.3 Cost savings - Potential Major Event

A potential major event occurs when there is an escalation in further assistance such as police and paramedics, which, again, was ultimately not needed. In this situation, without monitoring, an ambulance may be called by the client or the carers as they are unclear on a situation but are concerned enough to request a call out. If the event was a 'false alarm' the ambulance cost to the client or family member can be over $500. Ambulance coverage in insurance policies often does not cover callouts when transportation was not needed, so even with insurance, the client can be left with the bill.

Alternately, with monitoring technology, family members will have more information to decide if emergency services should be called or if they should check on the family member themselves. Due to the higher costs involved in such an event, the return on investment calculation may only need to need to occur 1-2 times a year to cover costs. Again, If the client is known to be 'at-risk' of such events, then the provider can further outline the benefits of the system to them.

## 4.2 Faster response to emergencies

A rapid response when a critical incident does occur can have a significant effect on clinical outcomes and quality of life. Critical incidents like falls, heart attack or a stroke, require an immediate response as time plays a vital role in the duration of recovery time and consequent support requirements.

Falls are potentially life-threatening incidents for the elderly, primarily if those falls result in injuries such as broken hips. There is the potential for significant disruption to the individual's health, the negative impact on family members and the economic costs of that care as part of rehabilitation to recover from a fall or other injury. The consumer that falls and cannot contact assistance due to immobility is likely to require significantly more care and rehabilitation than someone whose fall was detected and was attended to quickly.

The system can also trigger emergency services along with notifying family. This automation can ensure more immediate assistance in situations where the fall is more serious, and the client struggles to respond. Devices such as wearables with push-button alarms are not always utilised in emergencies, for example, if the individual is unconscious. The benefits of a passive smart home system that does not require activation from the user in such circumstances are clear. The implications for aged care providers are that timely responses and a reduction in unattended falls may negate extra strain on funding packages allowing for the provision of more targeted preventative services.

Given the implications for the broader community concerning the impact on hospital and medical services, the impact on lost productivity of family members taking time off to look after loved ones as well as impacts for the service provider; the costs to the community are high. While this scenario has focused on falls, many health conditions will benefit from a rapid-response.

**4.3 Trend Analysis**

Data collected over time can be used as a benchmark to identify when significant changes are beginning to occur. Behavioural changes can more easily be identified, and more specific care supports applied to better manage risk and develop capacity building and or wellness programs.

Trend analysis offers a service provider the ability to intervene earlier with regards to conditions that may deteriorate slowly over time, such as Mild Cognitive Impairment (a precursor to dementia). Similar to a rapid response scenario whereby an early intervention can mean fewer complications, trend analysis can do the same, but with conditions where changes may be more subtle and over long periods. Using A.I., it can identify behaviours that the carer or client may not necessarily notice such as a gradual decline in movement. Where declines occur more slowly and are not readily identified by untrained people, they can be identified with trend analysis software. This change then prompts an opportunity to discuss what might be occurring with the client - to determine if this is part of a more permanent deterioration or where acute care can be provided earlier.

Trend analysis offers a better prediction of when service requirements may increase, and further funding sought; this can help to minimise scenarios whereby the level of funding is not sufficient for the individual. Applications for additional funding can also have a more robust evidence base as data from the system could be used. It also offers the ability to better plan staffing requirements and time allocation. Trending of behaviours can also be positive as a client recovers from illness or injury. By seeing which clients are trending downwards and those trending upwards, staffing allocations may be made more accurate in servicing client needs, directed to where it is needed most.

These savings can be further increased by the avoidance of critical events and deploying capacity building activities to reduce the 'normal' rate of deterioration in a client's capabilities. Smart home systems (SHS) then become a data collection tool which aged care providers can discuss with the client if more specific services may be required, and if additional value-added services might be beneficial. The client may be more willing to pay if they can see a clear benefit.

At a higher level, trend analysis can be combined from multiple clients to understand better the patterns of behaviour and the necessary level of support for larger population cohorts. This 'big data' would assist in identifying patterns in local population health initiatives, many of which are known to be context-sensitive and need local data to adjust service provisioning. Technology-based measurements which are more objective could also be used as part of government calculators, requiring evidence from community implementations of such technology to establish the level of services in funding needed.

**4.4 Tailored Staff Allocation**

As previously mentioned, with better monitoring and data, a more tailored approach to care can be discussed with the client and then applied. The consequence of having access to richer client data is the capability for timelier or 'just-in-time' staff training and more finessed workforce allocation. This scenario considers the economic impact of implementing a more detailed model of staff resources directly in line with client needs. This data can provide more tailored and therefore, effective services, the objective of which is to increase staff productivity and client outcomes. At present staffing allocation of services to the client are generally conducted at routine intervals, with more intensive care applied when an issue becomes significant enough to be noticed. Instead, with the data provided by such systems, visits can be more targeted, increasing when they are needed more, and less so when they are not as necessary.

As one of the most significant line-item expenses in aged care provision, staffing productivity improvements and staff savings is a key area for investigation. By having a range of employees to call on that have broadened their core skills base with low-cost short courses could mean that more effective care can be achieved from a more traditional 'personal care worker' visit. While there might be a slight cost increase with the session itself, the carer is providing a higher level of service in that one visit so that sending another much higher cost (e.g. clinical level) staff member out at a different time may not be required. That has the potential to then lessen the overall number of visits needed. The cost benefits of a lower number of visits are not merely just the wage savings of that visit, but also include additional travel time and case management.

Upskilling of personal care workers through joint training sessions and education provides an opportunity for personal care workers to participate more effectively as part of a multidisciplinary team, achieving improved customer outcomes and often greater job satisfaction for the worker. With staff members, broadening their skillset more specifically in-line with client needs, effective teamwork and goal-directed care can be achieved. In the ageing population, where multiple complex conditions often present, hospital admissions can often be prevented with targeted multidisciplinary team approaches.

**4.5 Broader Value-added benefits**

In the course of the financial modelling and preparation of this report, the following additional value-added benefits have been identified. They were not included in the financial modelling but are listed as additional benefits with the potential to derive further economic impact. They may provide additional opportunities for income generation or further savings in expenses. These areas are suitable targets for economic analysis.

### 4.5.1 Building client compliance

Medical support is often limited to client compliance. Doctors and healthcare professionals can provide patients with the medication and strategies to improve their health but if they don't comply with taking medication or following through with instructions given by healthcare providers their health will suffer, and in the longer term this has the potential to result in much higher healthcare costs due to more severe outcomes or complications emerging.

There is a significant opportunity here for aged care providers to become client "health coaches", becoming more of a co-design partner and social-emotional coach in educating and helping customers develop more self-agency in their own care and support. CDC requires greater involvement of the client in their healthcare. Features in SHS such as medication alerts, reminders to put on wearable devices before leaving the home, and customised reminders to exercise, to attend appointments and to drink water can all be ways to increase their engagement in their own health goals and therefore achieve better quality of life and health outcomes.

### 4.5.2 Relationship Building

The management of the patients healthcare involves much more than simply instructing the customer what to do and them complying. There needs to be active engagement by clients in their own wellness protocols as part of their own treatment plan. This can be achieved more effectively if they play a more participatory role in the discussions around the benefits of their healthcare plan. Data from a smart home system can be used as discussion points for changing behaviour – both in capacity building and decline. For example, if a client is given instructions such as exercise and is then non-compliant according to the data, it can be a prompt for discussion to tease out why they are not complying and to help change their motivation, the exercise or instructions to better suit the client. It can also be used as conversation pointers for clients that say they are exercising when the data clearly shows otherwise. Discussions can be had with the client to delve further into their reasoning to then to help motivate or coach them to be more in train with their own goals.

### 4.5.3 Case studies and client referrals

The report from the National Ageing Research Institute showed that clients had positive attitudes towards this particular smart home system. As more systems are installed, and an increasing volume of people use them, there will likely be a growing number of case studies where its effectiveness in critical incidents and identifying early trends to better provide healthcare. These case studies could be used to improve aged care support services and processes. Codesigned case studies will increase the numbers of clients in terms of demonstrating operational effectiveness but more importantly ensuring greater peace of mind. The clients themselves talking to friends and family also increase the word-of-mouth benefits of the system. If aged care providers are seen as having expertise in new technology, and trusted smart home services, in particular, this can attract future clients.

### 4.5.4 Client piece of mind

Living alone and feeling isolated can become stressful for many individuals. Part of this stress can come from the potential ramifications of a critical incident and no one around to assist them. Stress can negatively impact health and can slow recovery from illness. By having a system that is unobtrusively monitoring them in case of emergencies and critical incidents, that can automatically alert friends and family; this can give the client the peace of mind to lower their stress levels. This is a major factor in the quality of life.

### 4.5.5 Staff compliance and dealing with complaints

The smart home system can assist in the quality management of staff in providing their duties to clients. By monitoring door sensors, providers can check what time a staff member enters and leaves a premises in conjunction with other monitoring such as GPS. This has the added benefit of quickly resolving potential disputes between clients and staff. For example, if a client makes a complaint that a staff member did not spend the entire time required at the client's home, and the staff member instead left early and spent the remainder of the time sitting in their car, the system could identify what time the staff member entered and left the premises by the door sensors. Dealing with even minor complaints can become quite time-consuming and can put additional administrative strains on a business. With better monitoring, this is much more easily resolved.

## 5. Discussion and conclusion

In summary, SHS can provide unobtrusive and objective data collection. A range of sensors placed throughout the home can collect vast amounts of data which can be summarised and analysed using artificial intelligence technology. Sensors can continually monitor and feed information to AI based system that can then advise of critical incidents, including those that may require emergency services, the intervention of the provider or the client's family and friends. This same data over time can show clear trends and identify changes to those trends, to which the client and their carers may or may not be initially aware. This trend analysis of behavioural changes can be further utilised to pinpoint where early intervention strategies might assist. Alternately the same trend analysis can be used when treatments are applied to show improvements in behaviour, by checking behaviours before and after treatment, further verifying the effectiveness of those treatments. Importantly, if treatments are not effective, this can be identified early, and care strategies can be updated. Please see figure 2.

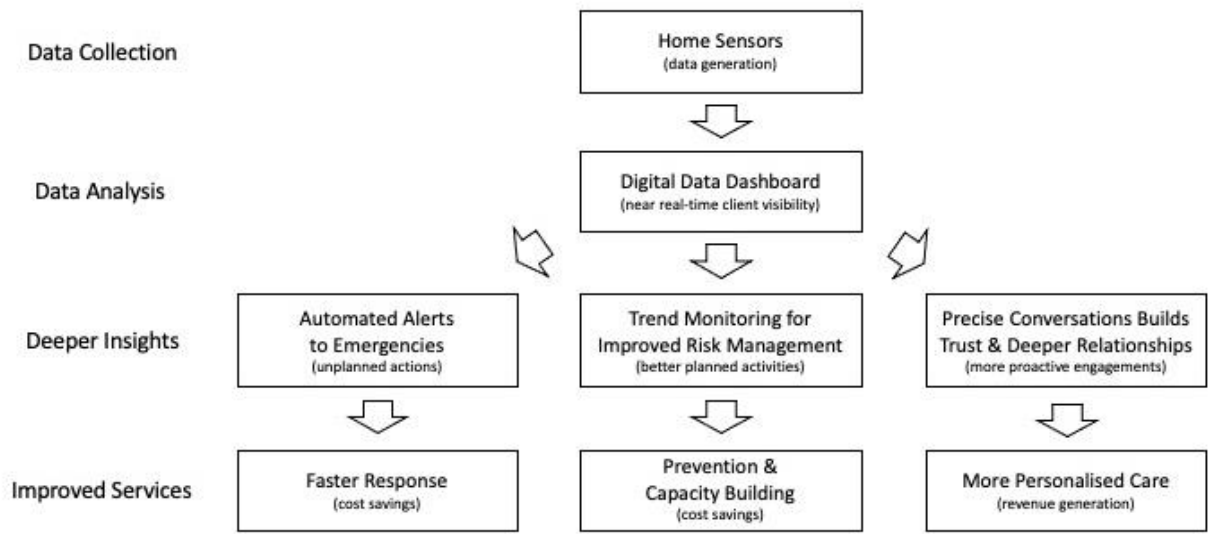

*Figure 2: Summary of the main economic benefits*

5.1 The Benefits for our Aged Population and their Families

As highlighted in the scenarios, there are both peace-of-mind benefits to our aged population and their carers. Merely knowing that if an adverse event were to occur and the client was alone, that the system could intervene and alert emergency response without the need for intervention by the client. The client and their carers ultimately determine the economic value of such a feature, but, it does become a point of discussion and greater engagement with the service provider.

Discussion points based on data generated from such systems foster deeper relationships with the customer as well as providing better health outcomes by actively adressing identified issues earlier.

5.2 The National Perspective

SHS may assist on a national level in a number of ways. As a monitoring tool, such systems can support individuals to stay-at-home longer and lessen the burden on residential aged care, which has a much higher economic cost to the Government. It is also clear that for the vast majority of the elderly, the preferred option is for them to live in their own home for as long as possible. Home monitoring systems can assist in achieving this via 24-hour monitoring for emergency/critical incidents and changes in behavioural trends.

Of further significance to Government is the more objective measures that technology can provide.

One of the main challenges for governments is to provide both equitable care for all citizens and the right level of care for each individual under a budgetary constraint. While this is being addressed via more nuanced levels of funding and tools to help determine those levels, subjective interpretations still form a major component of the application of those tools. Objective big data collected over more extended periods of time can show a much more accurate picture of an individual's particular situation.

While scientific randomised controlled trials applicable to the masses are the foundation of that evidence base, there are opportunities to develop more flexible care programs for our seniors tailored more specifically to their individual needs. For example, there are many scientific treatments for back pain; these might include medication or physiotherapy. A monitoring system can determine which is working best (or not working) for that individual.

Also, as has been shown as part of this study, there are opportunities to identify behaviours that are gradually changing or have the potential to cause a critical incident and to apply treatments or interventions before they deteriorate further. Then later providing rapid feedback as to if or to what degree the interventions have been effective via monitoring changes in behaviour. Again, this can be done in an objective and unobtrusive way that can easily show the true value of that intervention. This methodology follows the natural progression towards value-based healthcare.

Furthermore, there is the opportunity to combine data from individuals in a de-identified way to provide additional research into new treatments or interventions and help direct where resources are needed most, and which are more effective; this will accelerate healthcare research.

Two areas of caution should be noted. One is the safety of such information. Any online information has the risk of being hacked. This risk needs to be considered against the potential benefits of the collection and storing of such information. While there are examples of the general acceptance of sensitive information being stored and sent online (for example banking), healthcare information being sent and accessed online is an area that is still in the process of gaining acceptance by the community. The second area of caution is the scale of this study. While financial modelling is based on robust principles, the scenario analysis is based on a small number of smart home implementations; this is, in effect, a small-scale study. A much larger cohort of participants is needed to reinforce these findings.

## 5.3 The Benefits to Service Providers

The economic modelling demonstrates that deploying SHS can offer significant economic benefits to age care providers. However, before any technology is implemented, particularly in people's homes, it is recommended that clients be appraised for their suitability. There are characteristics of clients which should be considered when advising potential customers as to the benefits of smart home systems. According to the financial modelling, this includes clients likely to better understand their own situational health risks by using this type of technology. These clients would be identified by the service provider as those that would show behaviours that are clearly indicating a risk of critical incidents or that are likely to show a significant deterioration in the not too distant future. The client's financial position and level of funding also need to be included as part of these discussions. The previously outlined scenarios and their returns on investment are suitable case studies in how to position this with clients before requesting to purchase and install a smart home monitoring system.

The other clear category of potential clients is family members or carers that are more concerned with having peace of mind rather than any economic implications of the system. Levels of apprehension in the family are sometimes underappreciated and support around the issues these causes are not often provided. In these cases, the client or their loved ones may cover the cost of the system and its running costs simply to enhance their peace-of-mind.

More specifically, for aged care providers as a business, there are several economic and value-added benefits that will have a positive impact, both from an increase in revenue and to drive cost savings. SHS can improve current service strategies by providing greater visibility into day-to-day client needs and thereby offering the opportunity to develop more personally tailored services with their clients and then more closely tracking the improvement in the quality of services provided. This gives the business opportunities to offer more value-added services because the client and their carers can understand the benefits to them.

There are also opportunities to include new service strategies such as improved client risk management via trend analysis whereby emerging health risks can be discussed with the client, explaining the risks in those emerging trends and deciding on 'wellness' goals and a suitable course of action with the client. Capacity building programs that mitigate that risk and improve quality of life can then be introduced. Once these programs are introduced, they then can be measured over time, and clients can review the results.

More client specific insights and prevention strategies can also be discussed with families. At a national level, this aggregated data can begin to be used for input into industry regulation and reporting purposes. This type of technology provides aged care providers with much more detailed insights into each client. These insights allow them to develop trust and deeper relationships as part of the usual schedule of care. As a consequence of this, the client's journey from home care into residential aged care can also be better managed; this is a difficult transition to make – both psychologically and financially. Ideally, for many people, this could be a gentler transition than is often the case today.

With the supporting evidence of behavioural changes over time, aged care providers have the potential to be a more integrated part of each client's lifecycle from a basic home care package to residential aged care. By applying much more personally tailored services as they are needed from an objective behavioural change evidence base, the provider can become a more integrated part of the client journey. This evidence can assist aged care providers in applying for further funding if the client has a demonstrated need for it.

5.4 Conclusion

In conclusion, health interventions must include economic analysis; this is especially true of technologies that utilise big data, whereby the greater the level of uptake, the greater the amount of data that can be generated, managed and therefore be utilised. That high level of uptake is impossible if it is not economically viable.

Based on the scenario findings of the pilot study of an in-home monitoring system, the secondary economic analysis of those scenarios show that there are economic benefits along with other valued added benefits in the implementation of such systems. These include the ability to deliver more targeted and personalised care based on big data which is analysed via AI. This more personalised care shifts health delivery from a reactive model to one which is preventative and more actively case managed. Client-centric healthcare is seen in the Australian Aged Care system and the National Disability Insurance Scheme as better in managing chronic and complex conditions. In-home monitoring such as this fit within those models very effectively. It can help foster deeper relationships with the customer, and their carers as data can be used as discussion points highlighting potential issues to the client and ultimately assisting in developing new value-based healthcare models. In this case a big data perspective (Miah, Camilleri and Vu, 2021) can be adopted possibley for a big data analytics design (Miah, Vu and Gammack, 2019a and 2019b; Miah, Vu, Gammack, McGrath, 2017) through which various heathcare smart solutions would be developed and tested in problem context, adopting design scince research methodologies (e.g. Miah, 2008; Miah, Gammack and McKay, 2019; Genemo, Miah and McAndrew, 2015)

References

Aged Care Sector Committee. (2016). Aged Care Roadmap. Retrieved from

Amiribesheli, M., & Bouchachia, H. (2018). A tailored smart home for dementia care. Journal of Ambient Intelligence and Humanized Computing, 9(6), 1755. doi:10.1007/s12652-017-0645-7

Arnott, D., Lizama, F. & Song, Y. (2017). Patterns of business intelligence systems use in organisations, Decision Support Systems, 97, 58–68

Arrieta, A., Woods, J. R., Qiao, N., & Jay, S. J. (2014). Cost–benefit analysis of home blood pressure monitoring in hypertension diagnosis and treatment: an insurer perspective. Hypertension, 64(4), 891-896.

Australian Government Department of Health. (2019, 10th March 2020). National Aged Care Mandatory Quality Indicator Program. Retrieved from https://www.health.gov.au/initiatives-and-programs/national-aged-care-mandatory-quality-indicator-program

Campos, F. d. B. V. d. C. (2016). The impact of user-generated content on Facebook advertising performance.

da Costa Campos, F. D. B.V. (2015). The Impact of User-generated Content on Facebook Advertising Performance, Published Dissertation, CATÓLICA LISBON School of Business & Economics, UNIVERSIDADE CATOLICA PORTUGUESA, (https://repositorio.ucp.pt/bitstream/10400.14/20260/1/Dissertation_Francisco%20Campos_final.pdf), (accessed on 20 March 2018)

Daniel, J.-Q., Irving, A. C.-A., Veronica, M. G.-S., & Luis, A. M.-H. (2020). Smart Sensor Based on Biofeedback to Measure Child Relaxation in Out-of-Home Care. Sensors, 20(4194), 4194-4194. doi:10.3390/s20154194

Genemo, H., Miah, S.J., and McAndrew, A. (2015). A Design Science Research Methodology for developing a Computer-Aided Assessment Approach using Method Marking Concept, Education and Information Technologies, 21, 1769–1784

Gill, L., & Cameron, I. D. (2020). Identifying baby boomer service expectations for future aged care community services in Australia. Health & social care in the community. doi:10.1111/hsc.13187

Lazarou, I., Karakostas, A., Stavropoulos, T. G., Tsompanidis, T., Meditskos, G., Kompatsiaris, I., & Tsolaki, M. (2016). A novel and intelligent home monitoring system for care support of elders with cognitive impairment. Journal of Alzheimer's Disease, 54(4), 1561-1591. doi:10.3233/JAD-160348

Miah, S.J. (2008). An ontology based design environment for rural decision support, Unpublished PhD Thesis, Griffith Business School, Griffith University, Australia

Miah, S.J., VU, HQ., Gammack, JG, and McGrath, GM. (2017). A Big-Data Analytics Method for Tourist Behaviour Analysis, Information and Management 54, 771-785

Miah, S.J., Gammack, JG, and McKay, J. (2019). A Metadesign Theory for Tailorable Decision Support, Journal of Association for Information Systems 20 (5), 570-603

Miah, S.J., Camilleri, E., and Vu, HQ (2021). Big Data in Healthcare Research: A survey study, Journal of Computer Information Systems, 32(3)

Miah, S.J., Vu, HQ, and Gammack, JG (2019a). A Big-Data Analytics Method for capturing visitor activities and flows: the case of an Island Country, Information Technology and Management 20 (4), 203-221

Miah, S.J., Vu, HQ, and Gammack, JG (2019b). A Location Analytics Method for the Utilization of Geo-tagged Photos in Travel Marketing Decision-Making, Journal of Information and Knowledge Management, 18(1), 1950004

Piggott, J., Kendig, H., & McDonald, P. (2016). Population Ageing and Australia's Future: ANU Press.

Rahman, A., Akbar, D., Rolfe, J., & Nguyen, J. (2019). Developing a population wide cost estimating framework and methods for technological intervention enabling ageing in place: An Australian case. PLoS ONE, 14(6), 1-19. doi:10.1371/journal.pone.0218448

Royal Commission into Aged Care Quality and Safety. (2019). Royal Commission into Aged Care Quality and Safety interim Report: Neglect. Retrieved from https://agedcare.royalcommission.gov.au/publications/Pages/interim-report.aspx

StewartBrown. (2020). StewartBrown Aged Care Financial Performance March 2020 Survey Sector Report. Retrieved from https://www.stewartbrown.com.au/news-articles/26-aged-care

Tarrant, A. (2016). Getting out of the swamp? Methodological reflections on using qualitative secondary analysis to develop research design, International Journal of Social Research Methodology, 20(6), 599-611.

Taylor, D., Barrie, H., Lange, J., Thompson, M. Q., Theou, O., & Visvanathan, R. (2019). Geospatial modelling of the prevalence and changing distribution of frailty in Australia – 2011 to 2027. Experimental Gerontology, 123, 57-65. doi:10.1016/j.exger.2019.05.010

Thorne, S. (1994). Secondary analysis in qualitative research: Issues and implications, In J. Morse (ed.), Critical issues in qualitative research methods, Thousand Oaks, CA: Sage.

Vanus, J., Machacek, Z., Koziorek, J., Walendziuk, W., Kolar, V., & Jaron, Z. (2016). Advanced energy management system in Smart Home Care. International Journal of Applied Electromagnetics & Mechanics, 52(1-2), 517-524. doi:10.3233/JAE-162028